\documentstyle[prl,aps,twocolumn,epsf]{revtex}
\begin{document}
\title{Oscillatory behavior of a superradiating system
coupled to electron reservoirs} 
\author{T. Brandes\cite{newaddress}, J. Inoue, and A. Shimizu} 
\address{Institute of Physics, University of Tokyo,
3--8--1 Komaba, Tokyo 153--8902, Japan}
\draft
\date{\today}
\maketitle
\begin{abstract}
We investigate a superradiating system coupled to external reservoirs. Under 
conditions where electrons tunneling at a rate $T$ act like an 
electron pump,
we predict a novel phenomenon in the form of oscillations with a frequency 
$\omega \simeq \sqrt{2\Gamma T}$ that appear in the (photon) emission 
intensity, where
$\Gamma $ is the spontaneous decay rate of a single two-level system.
The effect, together with a 
strong enhancement of the superradiant peak, should be observable in 
semiconductor quantum wells  in
strong magnetic fields, or in quantum dot 
arrays. 
\end{abstract} 
Superradiance occurs in  the spontaneous coherent decay of an initially 
excited ensemble of $N$ two-level systems  which are interacting with a 
common photon field. The corresponding emission rate of photons is 
proportional to $N^{2}$ which is abnormally large when compared to the 
incoherent decay of $N$ independent systems. 
Furthermore, the emission is not exponentially in time 
but has the form of a very sudden peak on a short time 
scale $\sim 1/N$. The phenomenon was predicted by Dicke in a seminal paper 
\cite{Dic54} and observed for the first time 
\cite{Skretal73} in an optically pumped hydrogen fluoride gas. 
Since then, many investigations of 
superradiance concentrated on modifications through geometry effects and 
dephasing processes such as dipole-dipole interactions which had been neglected 
in the original Dicke paper \cite{Benedict}. Since superradiance 
intrinsically is a many-body problem, 
this also gave the possibility to study the concept of coherence 
and dephasing in a many-body context 
\cite{Tokihiro}. 

In the present Letter, we propose an extension of the original Dicke model 
that in principle opens the possibility for an external control of 
superradiance. The main idea is to realize the superradiant `active region' 
in a semiconductor and to couple it to external electron reservoirs through 
tunnel barriers, thus allowing for a varying electron number. 
We predict that for reservoir conditions which work like an 
electron pump,
the initial 
coherent superradiant peak can be strongly enhanced 
if the tunnel rate $T$ is high enough. 
Furthermore, we predict a novel phenomenon in the form of 
strong oscillations of the emitted light with a frequency $\omega $ that in 
good approximation is given by 
\begin{equation}
\label{frequency}
\omega \simeq \sqrt{2\Gamma T}
\end{equation}
for large $T > 2\Gamma $, where
$\Gamma $ is the spontaneous decay rate of a single two-level system.
For smaller tunnel rates $T$, there is a smooth crossover to the conventional
Dicke peak \cite{Dic54} in the limit $T\to 0$.
In contrast to oscillatory superradiance in atomic 
systems \cite{Andreev}, these oscillations are not due to reabsorption of 
photons, but due to tunneling  of electrons into an active region which 
is described by  many-body wave functions. The latter are characterized by a 
total pseudo spin $J$ and a pseudo spin projection $M$ for $N$ electrons 
occupying upper and/or lower levels, including empty levels.
In contrast to the original Dicke problem, $J$ is no longer 
conserved but develops a dynamics that is driven by the tunneling process 
which leads to a coupling of $J$ and $M$, whose equations of motion have
oscillatory solutions.

We propose a concrete realization of the model in a semiconductor quantum 
well  in 
a strong magnetic field  which  defines degenerate energy levels 
for electrons and  holes. The `active layer' is driven in analogy to a 
semiconductor diode laser by injecting conduction band electrons 
and valence band holes that can radiatively recombine. 
In contrast to the diode laser, no stimulated emission processes are required. 

We first describe our model Hamiltonian. A photon field 
$H_{p}= \sum_{{\bf Q}}\Omega_{{\bf Q}}a^{\dagger}_{{\bf Q}}a_{{\bf Q}}$
with creation 
operator $a^{\dagger}_{{\bf Q}}$ for a mode ${\bf Q}$ gives rise to 
transitions which change  an internal degree of freedom $\sigma=(\uparrow, 
\downarrow) $ of one-particle electronic states labeled $(i,\sigma )$ with creation
operator $c^{\dagger}_{i,\sigma}$. The states 
are degenerate with respect to $i$ with energies $\varepsilon_{i 
\uparrow}=-\varepsilon _{i\downarrow}=\hbar\omega _{0}/2$. The 
electron--photon coupling matrix element $g_{{\bf Q}}$ is assumed to be 
independent of the electronic quantum numbers $(i,\sigma )$. This condition 
is fulfilled if the momentum matrix element $\langle i\sigma| {\bf p}| i' 
\sigma'\rangle \propto \delta _{ii'}$ for $ \sigma \ne \sigma'$ 
as is the case for the applications discussed below. 
Then, the Hamiltonian without reservoir 
coupling can be written as 
\begin{equation}
\label{dhamiltonian}
H_{D}=\hbar\omega _{0}\hat{J}_{z}+
\sum_{{\bf Q}}
g_{{\bf Q}}\left(a^{\dagger}_{{\bf Q}}+a_{{\bf Q}}\right) 
\left(\hat{J}_{+}+\hat{J}_{-}\right)+ H_{p},
\end{equation}
where the operators   
$\hat{J}_+:=\sum_i c^{\dagger}_{i,\uparrow }c_{i,\downarrow }$,
$\hat{J}_-:=\sum_i c^{\dagger}_{i,\downarrow }c_{i,\uparrow }$
and
$\hat{J}_z:=\frac{1}{2}\sum_i \left(c^{\dagger}_{i,\uparrow }c_{i,\uparrow }
-c^{\dagger}_{i,\downarrow }c_{i,\downarrow }\right)$
form a (pseudo) spin algebra with angular momentum commutation relations. 
Note that so far we have not refered to the explicit 
form of the one-particle 
states; the Hamiltonian Eq.~(\ref{dhamiltonian}) represents a whole class of 
physical systems rather than one specific experimental situation. The model 
and the examples below refer to a photon field; we point out, however, that 
our results are valid for a coupling to an arbitrary boson field 
(e.g. phonons), too.  

We allow the number of electrons $N$ in the active region as described by 
$H_{D}$ to vary by tunneling to and from electron reservoirs $\alpha= I/O$ 
(`In'and `Out') with Hamiltonians 
$
H_{\alpha }=\sum_{k}\varepsilon _{k}^{\alpha }
c^{\dagger}_{k,\alpha }c_{k,\alpha }
$
for non-interacting electrons 
with equilibrium Fermi distributions $f_{\alpha 
}$.
The tunneling of electrons is described by the usual tunnel Hamiltonian
$
H_{T}=\sum_{ki\sigma\alpha}\left( t^{\alpha }_{ki\sigma } 
c^{\dagger}_{k,\alpha }c_{i,\sigma  } + c.c. \right)
$
with coefficients $t^{\alpha }_{ki\sigma }$ that in general depend 
on the specific form of the tunnel barriers in real space and the
one-particle wave functions. 
The total Hamiltonian is given by the sum $H=H_{D}+\sum_{\alpha}
H_{\alpha }+H_{T}$. 
We note that in this Letter we do not consider explicitly 
the effects of 
electron-electron and electron-hole interactions but 
concentrate on the effects which result 
from the tunneling processes only.  
We do not expect the interactions 
	to lead to qualitative differences, 
	because in our case the carrier density 
	is high  
	and drastically varies 
	as a function of time, which weakens and smears out 
	the interaction effects. 

The coupling of the active region to reservoirs combines the dynamics of 
a (many) spin-boson problem, $H_{D}$, with the transport through it.
We use the master equation approach to calculate the dynamical 
evolution of observables like the emission intensity.
This method, while only perturbative in the coupling matrix elements 
$g_{{\bf Q}}$ and $t^{\alpha }$, has turned out to yield results that
seem to be in good agreement with experiments both for the superradiant 
problem \cite{Benedict,Andreev} and transport through small regions of 
interacting electrons (quantum dots) \cite{Weinmann,Grabert}. We note that 
correlations between different tunneling processes (co-tunneling) are 
neglected here.

The eigenstates of the active region are characterized by the total pseudo 
spin $J$ and its projection $M$ through  
$\hat{J}^{2}  | JM;\{\lambda\} 
\rangle=  J(J+1)   | JM;\{\lambda\} \rangle$ 
and $\hat{J}_z  | JM;\{\lambda\} 
\rangle=  M   | JM;\{\lambda\} \rangle$, 
where $\hat{J}$ is the total pseudo 
spin operator. Here, $\{\lambda\} $ denotes all additional quantum numbers 
apart from $J$ and $M$ that are necessary to characterize the eigenstates of 
$H_D$. Radiative transitions obey the selection rule $M\to M\pm 1$ with a 
spontaneous emission intensity $I_{JM}$,
\begin{equation} \label{emission} I_{JM}= \hbar 
\omega_{0} \Gamma \nu_{JM},\quad \nu_{JM}:= (J+M) (J-M+1). 
\end{equation}
Here, $\Gamma$ is the spontaneous emission rate of 
one {\em single} two-level 
system; for radiative transition in atoms, $1/\Gamma $ is in the nano second 
range. 

Furthermore, the rates $\Gamma_{JM\to 
J'M'}$ for transitions between eigenstates of $H_{D}$ 
through electron tunneling  may be written in the form 
\begin{eqnarray}
\label{transitionrate}
&\Gamma&_{JM\to J'M'}=\sum_{\alpha}
T^{\alpha}\left\{
\gamma _{JM\to J'M'}
f_{\alpha }(\Delta E)\right. \nonumber\\ 
&+& \left.
\gamma _{J'M'\to JM}
\left(1-f_{\alpha }(-\Delta E)\right)\right\}.
\end{eqnarray}
Here,
$
T^{\alpha }:=(2\pi/\hbar)\sum_{k}t^{\alpha }_{ki\sigma } (t^{\alpha 
}_{ki'\sigma' })^{*}\delta (\Delta E-\varepsilon _{k}^{\alpha }) $ 
is the tunnel rate for lead $\alpha $ which in fact is a  matrix in  
$(i,\sigma )$ and depends on the energy 
difference $ \Delta E$ between final and initial state. 
We neglect the energy 
and site dependence, 
furthermore the dependence on $\sigma $ is absorbed into 
the index $\alpha $ through the boundary conditions: the coupling  
to the reservoirs has the effect of a `pseudo spin-up pump' 
where only $\sigma 
=\uparrow$ electrons can tunnel in and $\sigma =\downarrow$ electrons tunnel 
out. That is, the chemical potential for $\alpha =I$ is assumed to be 
situated above all possible energy differences of states $| JM;\{\lambda\} 
\rangle$ with $M$ differing by plus $1/2$, and 
the chemical potential for $\alpha =O$ is chosen such that 
`down' electrons can tunnel out to $\alpha =O$, but not tunnel in. 
Furthermore, tunnel matrix elements for processes like tunneling from
$\alpha =I$ with $\sigma =\downarrow$, leading to $M'=M-1/2$, are assumed
vanishingly small; this condition is satisfied in the applications below.

The coefficients $\gamma$ in 
Eq.~(\ref{transitionrate}) are determined by 
the Clebsch-Gordan coefficients 
for adding or removing a single pseudo-up or down spin to 
the active region, 
$
\gamma _{JM\to J'M'}:=
|\sum_{m=\pm 1/2 }\langle J'M'| JM;j=1/2,m\rangle|^{2}. 
$
Here, we adopted the same approximation that 
has been used in \cite{Weinmann} 
for calculating matrix elements for transition rates in transport through 
correlated quantum dots: the specific form of the 
many-particle wave function 
in the active region (i.e. the quantum  numbers $\{\lambda \}$) 
is neglected, 
and the matrix elements are approximated  by the Clebsch-Gordan coefficients. 

\begin{figure}[ht]
\unitlength1cm
\begin{picture}(8,6)
\epsfxsize=6cm
\put(0.5,0.5){\epsfbox{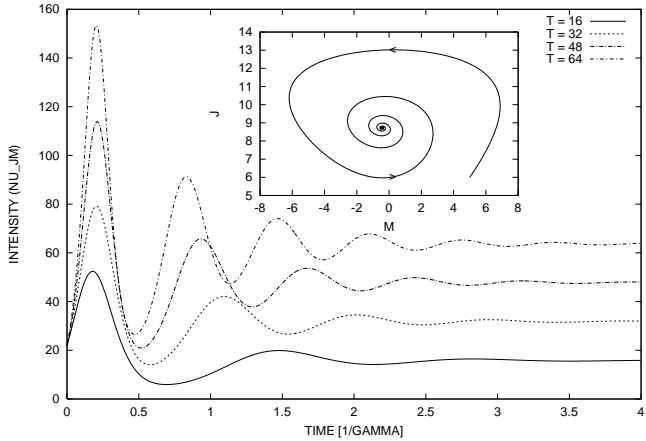}}
\end{picture}
\caption[]{\label{figure1}
Time evolution of the emission intensity 
$\nu_{JM}=I_{JM}/\Gamma \hbar\omega_{0}$ 
for different transmission rates $T$. 
Inset: $\langle J \rangle_{t}$ vs. $\langle M \rangle_{t}$ 
for $T=64$.} \end{figure}

We describe the dynamics of the active region in terms of probabilities 
$\rho (JM)_{t}$ which are the diagonal elements of the reduced density 
operator at time $t$ in the basis of eigenstates 
$| JM \{\lambda\}  \rangle$, 
where all information apart from the quantum numbers $J$ and $M$ has been 
traced out. In lowest order perturbation theory, there is no interference 
between $H_{ep}$ and $H_{T}$, and the master equation in Markov and Born 
approximation reads \cite{firstline}
\begin{eqnarray}
\label{masterequation}
&\dot{\rho}& (JM)_{t}=
-\Gamma\left\{
\nu_{JM}\rho (JM)_{t}-\nu_{JM\!+\!1}\rho(JM+1)_{t}\right\}\nonumber\\
&+&\sum_{J'M'}\left\{ \Gamma _{J'M'\to JM}\rho (J'M')_{t}
-\Gamma _{JM\to J'M'}\rho (JM)_{t}\right\}.
\end{eqnarray}
The time dependence of the expectation value of the emission rate, 
Eq.~(\ref{emission}), was obtained from the time evolution of $\rho 
(JM)_{t}$ by numerical solution of Eq.~(\ref{masterequation}). The result is 
shown in Fig. (\ref{figure1}) for an initially excited state with $J=6$, 
$M=5$. The initial emission maximum is the original `Dicke-peak', 
followed by 
oscillations that die out at an intensity proportional 
to the tunnel rate $T$ 
that was chosen symmetric, $T=T^{I}=T^{O}$. 
The frequency of the oscillations 
increases with $\sqrt{T}$ and follows in good approximation the law 
Eq.~(\ref{frequency}). In particular, the initial peak is strongly enhanced 
with increasing tunnel rate. This behavior is related to an initial increase 
of the total pseudo spin as can be seen from the `phase-space' 
plot (inset of 
Fig. (\ref{figure1})) of the expectation values of 
$J$ and $M$ which both oscillate in time.  

These observations can be understood as follows.  
In the original Dicke problem, a fixed total 
number $N$ of electrons with exactly one pseudo spin $\sigma $ on each site 
$i$ is assumed. Here, we work in a grand-canonical ensemble where $N$ varies 
through single electron tunneling:  doubly occupied or empty single particle 
levels $i$ become possible, i.e. there is no longer an exact `half-filling'. 
One immediate consequence is that a tunneling electron changes the quantum 
numbers $J$ and $M$. The 
change $\dot{J}$ of $J$ is proportional to $M$ itself, 
$\dot{J}(t) = T M(t)/J(t)$,
which follows considering the Clebsch-Gordan coefficients for 
adding a pseudo up spin.
At the same time, $M$ is 
increased by $1/2$ as is the case for out--tunneling of a pseudo down spin:
$M$ increases by $1$ at the tunnel rate $T$
and decreases by spontaneous emission at a rate $\Gamma \nu_{JM}$.
Therefore, $J$ and $M$ obey roughly
\begin{eqnarray}
\label{eom}
\dot{M}(t)&=&-\Gamma\nu_{J(t)M(t)}+T\nonumber\\ 
\dot{J}(t)&=& T \cdot M(t) /J(t),
\end{eqnarray}
which are governed by the two parameters $\Gamma $ and $T$, 
the emission rate and the tunnel rate.   
After eliminating $J$ in the equations of motion Eq.~(\ref{eom}), 
and approximating $\nu_{JM}\approx J^{2}-M^{2}$, we obtain
$
\ddot{M}- 2\Gamma
M\dot{M}+\omega ^{2}M=0
$
with $\omega = \sqrt{2\Gamma T}$, Eq.~(\ref{frequency}). 
For $T>2\Gamma$, 
this describes a harmonic oscillator  
with frequency Eq.~(\ref{frequency}) and amplitude dependent 
damping. For smaller $T$, the oscillations are no longer visible and 
Eq.~(\ref{frequency}) does no longer hold. For $T\to 0$, there is a smooth 
crossover to the conventional Dicke peak with vanishing intensity at large 
times and without oscillations. We again point out that our model 
applies to the small sample limit of the superradiant problem:
reabsorption processes of photons that may lead to oscillatory behavior of 
the intensity do not play any role here.

The oscillator equations Eq.~(\ref{eom}) in fact
are the quasiclassical limit of the  
master equation Eq.~(\ref{masterequation}) for $J\gg 1$. 
Neglecting fluctuations of the 
expectations values $M(t)$ and $J(t)$, one writes \cite{firstline} the 
time dependent probability distribution as $ \rho (JM)_{t}=\delta_{M,M(t)} 
\delta_{J,J(t)}$. 
The detailed form of the intensity peak and the intensity oscillations 
obtained in this way deviate from the exact solution of 
Eq.~(\ref{masterequation}), whereas the qualitative features 
remain unchanged 
which we checked numerically. In the limit of large $J$, one obtains the 
two equations Eq.~(\ref{eom}) for $J$ and $M$.        

\begin{figure}[ht]
\unitlength1cm
\begin{picture}(6,4.5)
\epsfxsize=5cm
\put(0.5,0.5){\epsfbox{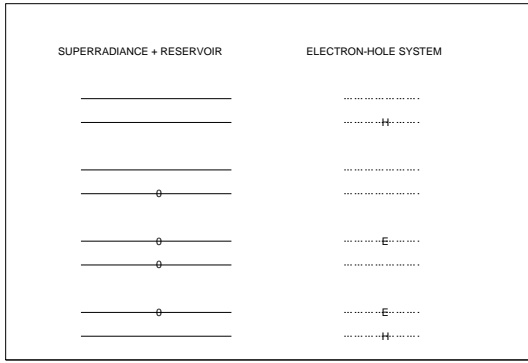}}
\end{picture}
\caption[]{\label{figure2} Correspondence of our model 
(superradiance with reservoir) with an electron-hole system. 'E' denotes an 
electron in the conduction band, 'H' a hole in the valence band.} 
\end{figure}

We now turn to the question in what physical systems
the effects described above can be 
observed experimentally. 
We note that the tunneling processes can be replaced with classical
injection processes over potential barriers, because we have assumed
that quantum correlation is absent between subsequent tunneling
processes. 
We thus propose the system of electrons and 
holes in semiconductor quantum wells in strong magnetic 
fields.  Vertical injection of conduction-band electrons and 
valence-band holes into an active region acts 
like the pumping mechanism 
described above. In fact, this mechanism is exactly what is used in
lasers or 
light emitting diodes with forward biased pn junctions. In our case, mirrors 
as in a laser are not necessary, in particular stimulated emission processes 
must play no role. 
The strong magnetic field is necessary to have dispersionless 
single electron 
levels $i=X$, corresponding to the lowest 
Landau bands ($n=0$) and guiding 
center $X$ \cite{LL3} in the conduction and the valence bands. 
In this case, the interband optical 
matrix elements are diagonal in $i$. 
The correspondence with our model can be 
seen by mapping its four basic single particle states to the states of the 
electron-hole system (Fig. (\ref{figure2})): the empty state becomes 
the hole (H), the pseudo-spin down electron becomes the empty state, the 
doubly occupied state becomes the electron (E), and the pseudo-spin up 
electron becomes the state with one electron and one hole. 
The number of total electrons $N$ in our model has its 
correspondence via 
\begin{equation}
\label{numbercorrespond}
N=N_E+N_{s}-N_H,
\end{equation}
where $N_{s}=\Phi/\Phi_0$ is the degeneracy for a given magnetic flux $\Phi$
($\Phi_0=hc/e$ is the flux quantum), $N_E$ the number of electrons in the 
conduction-band, and 
$N_H$ the number of holes in the valence-band. 

We predict that an initial optical 
or current excitation of the system leads to a 
superradiant peak of emitted light that becomes strongly  enhanced if the  
tunneling rate becomes higher. Furthermore, subsequent oscillations of the  
emitted light should be visible at an approximate frequency 
Eq.~(\ref{frequency}). We also expect similar oscillations to be visible as 
weak corrections to the injection current; 
detailed calculations will be presented elsewhere. 

As a second experimental setup, 
we propose an array of identical quantum dots, 
coupled to electron reservoirs as above. The array must have the capability 
to coherently radiate, where each dot has a pair of well-defined internal 
levels that allow for transitions under emission of photons. 
Another possible realization could be a geometry as in the  quantum cascade 
laser which has been proposed recently \cite{cascade} as an alternative to 
conventional semiconductor diode lasers. In this case, transitions between 
different electronic subbands lead to photon emission. 
Note that in all three 
cases, the photon escape time $\tau =L/c$ ($L$ is the linear dimension
of the active region and 
$c$ the speed of light) has to be much smaller than all other time scales of 
the problem, because in our model we assumed the `small sample 
superradiance case' where reabsorption effects play no role. 

We mention that all the effects described above should in principle
be observable not only for photons, but also for other bosonic fields such 
as phonons, or magnons.

Finally, we point out that so far 
we have not addressed the question of phase 
coherence. In fact, the conventional superradiance is a transient process 
that occurs only on a `mesocopic' time scale with an upper boundary 
\cite{lowerboundary}
given by a phase coherence time $\tau _{\phi}$. Inelastic 
processes such as dipole-dipole interactions \cite{Tokihiro} in general 
destroy the phase coherence between single particle states and the 
description using the Dicke states with well-defined $J$ and $M$ becomes 
void. On the other hand, coherence between states with different $J$ and 
different $M$ is not required in our formalism, on the contrary this would 
require consideration of coherent tunneling which is beyond the scope of our 
approach. We assume that dephasing  processes are weak such that the 
time to observe the initial Dicke peak and some cycles of the
subsequent oscillations is still shorter than $\tau _{\phi}$. 
An ideal case would be a strong initial excitation to a high initial 
pseudo spin $J$, and a high tunnel rate $T$ that yields fast oscillations.
Furthermore, strong magnetic fields in general suppress scattering rates 
although at the present state we can give 
no quantitative estimates for $\tau_{\phi}$.

This work has been supported by the 
Core Research for Evolutional Science and Technology (CREST), JST.


\begin{thebibliography}{10}
\bibitem{lowerboundary}
{The lower boundary is given by the photon escape time $\tau$.}
\end{thebibliography}

\end{document}